\documentclass[8pt,twocolumn]{article}
\usepackage{cite}
\usepackage{graphicx}
\usepackage{amsmath}
\usepackage{bm}
\usepackage{url}

\newcommand{\iset}{{\{i\}}}
\newcommand{\jset}{{\{j\}}}
\newcommand{\ijset}{{\{i,j\}}}

\begin{document}

\begin{titlepage}
  \quad\\[1cm]
  \makeatother
    {\Huge IEEE Copyright Notice}\\[0.5cm]
    {\large This work has been submitted to the IEEE for possible publication. 
    Copyright may be transferred without notice, 
    after which this version may no longer be accessible. }
\end{titlepage}

\title{Copula-Based Density Estimation Models for \\ Multivariate Zero-Inflated Continuous Data}
\author{Keita~Hamamoto$^1$\thanks{keita.hamamoto.uv@hitachi.com}}
\date{%
    $^1$Central Research Laboratory, Hitachi, Ltd., Kokubunji 185-8601, Japan\\%
}

\maketitle

\begin{abstract}
  Zero-inflated continuous data ubiquitously appear in many fields,
  in which lots of exactly zero-valued data are observed while others distribute continuously.
  Due to the mixed structure of discreteness and continuity in its distribution,
  statistical analysis is challenging especially for multivariate case.
  In this paper, we propose two copula-based density estimation models that 
  can cope with multivariate correlation among zero-inflated continuous variables.
  In order to overcome the difficulty in the use of copulas due to the tied-data problem in zero-inflated data, 
  we propose a new type of copula, rectified Gaussian copula, and present efficient methods 
  for parameter estimation and likelihood computation.
  Numerical experiments demonstrates the superiority of our proposals compared to conventional density estimation methods.
\end{abstract}

\section{Introduction}\label{sec:introduction}
Density estimation is one of the most important tasks 
in many applications such as statistical machine learning, anomaly detection, and so on. 
The goal of the density estimation is to estimate the probability density function (PDF) from 
the observed data. 
Estimated PDF provides rich information of the properties of data including skewness, multimodality, cluster structure, correlations etc., 
which can be immediately used as a building block for subsequent analysis and applications.
Effective density estimation tools are already available such as 
multivariate Gauss model, Gaussian Mixture Model (GMM) and Kernel Density Estimation (KDE)
for multivariate continuous data~\cite{PRML,Parzen}.
Multivariate Discrete distributions are also handled for example by
a family of multivariate Poisson distributions~\cite{schulz21multivariatepoisson},
in which joint probability \textit{mass} function is estimated.

For mixed random variable, a mixture of continuous variable and discrete variable, 
the modeling of its distribution needs special care~\cite{pishro16introduction}.
Zero-inflated continuous data is a member of the mixed random variable,
which simultaneously has continuous observations 
and substantial amount of exactly zero-valued observations.
For zero-inflated \textit{discrete} variable, 
there are well known parametric models for its distribution
such as zero-inflated Poisson~\cite{lambert92zip} and 
zero-inflated negative binomial distribution~\cite{ridout01zinb},
whereas zero-inflation sometimes happens in continuous variable. 
Variables like drug dosage, alcohol consumption and 
amount of certain biomarker in biomedical field~\cite{liu19statistical}, and
amount of outstanding loan, spouse's income, 
and collateral value in financial domain are typical examples
of zero-inflated continuous data. 
Those kind of zero-inflation is inevitable for many real-world applications
since they naturally arise for example when data are collected from multiple sources 
like administration/control groups or loan services for singles/married.

In one dimensional case, PDF of zero-inflated continuous variable is the form of
\begin{equation}\label{eq:pdf1dim}
f(x) = q\delta(x) + (1-q) \tilde{f}(x)
\end{equation}
where $q$ is the probability of zero occurrence, $\tilde{f}(x)$ is the PDF for nonzero data. 
Dirac's delta-function $\delta(x)$, whose value diverges at the origin and vanishes elsewhere,
is used to express the infinite density of point mass~\cite{pishro16introduction}.
We will focus on zero-inflated nonnegative continuous data in this paper 
and $\tilde{f}(x)$ takes finite value only for $x>0$.
The point \textit{mass} with diverging PDF at the origin 
and finite \textit{density} at positive domain have different 
probability measures and often result in a multimodal structure.
In multivariate case, the modeling of PDF becomes much more difficult. 
In $D$-variate zero-inflated data, each data point resides in one of $2^D$ subspaces.
Each of subspaces corresponds to a set of zero-valued variables out of $D$ variables and 
has different dimensionalities. 
For example for $D=3$, joint probability is the mixture of one point mass at the origin,
three line densities along each axis, three surface densities on plains spanned by two axes, and 
one volume density.
In this paper we use the term density for all types of those densities, sometimes including mass.

Those diverging PDF, exponentially large number of subspaces, and different dimensionality make the density estimation
of multivariate zero-inflated continuous data intractable. 
One possible approach is brute force subspace-wise density estimations. 
However, as the number of subspaces is exponentially large for high dimension, 
the computational cost becomes impractical.
Also, performance of subspace-wise density estimation degrades since the number of data points
residing in each subspace exponentially decreases in high dimension.
Conventional GMM, being able to handle the multimodal distributions, 
also fails to capture all the subspaces
unless one uses exponentially large number of Gaussian components. 
Moreover, even for a low dimensional case, parameter estimation often leads to infinitely narrow and steep Gaussians along axes of coordinate, 
which results in unfairly large likelihood value with poor fitness to the data.
KDE as well fails to capture the different dimensionality.
As the proper bandwidth of density kernel should strongly 
depends on the dimensionality of subspace they reside, 
the estimated PDF becomes unpleasantly broad toward direction perpendicular to each subspace
and assigns large PDF value to data-sparse regions.
In addition, as well-known, KDE performs poorly for high dimension 
due to so called "curse of dimensionality~\cite{scott15multivariate}.
So we need a new density estimation method which can properly handle multivariate zero-inflated continuous data.

In this paper, we propose two density estimation methods based on copula, each of which handles 
different mechanisms of zero inflation. We use mixed random variable model for 
marginal distribution as in eq.\eqref{eq:pdf1dim} 
and copula density to capture the multivariate correlations. 
Those multivariate mixed random variable modeling allows us to handle 
the different dimensionality of subspaces,
in which, for example, mass, linear density, surface density, and volume density
are all expressed in a unified way unlike conventional GMM and KDE.
We develop a new type of copula, rectified Gaussian copula, 
in the second model to avoid tied-data related difficulty in the 
application of copula to zero-inflated data.
We also show polynomial time algorithms to estimate parameters of models, and to compute likelihood values.
Numerical experiments for synthetic and real data show higher performance
of our methods compared to conventional GMM and KDE.

\section{Preliminaries}

In this section, we introduce some basic concepts and notations to better understand our proposals.
First, we briefly summarize the notations and usages of copula. 
We show that direct application of conventional copulas to zero-inflated data is not appropriate due to well-known tied-data problem.
Finally, we present two different mechanisms of zero inflation in association with the 
concepts developed in missing data analysis. 

\subsection{Copula}
Copula, a paradigm in statistics for multivariate modeling,
attracts growing attentions due to its powerful capability
and flexibility for the multivariate modelings~\cite{copulabook}.
Since its first proposal by Sklar and several subsequent developments~\cite{joe96families,bedford02vines,aas09pair},
copula has been applied in variety of fields such as finance~\cite{bouye00copulas,patton12review},
reliability analysis~\cite{fang20copula},
survival analysis\cite{georges01multivariate},
clinical medicine~\cite{burzykowski01validation}, and so on.
In the formulation using copula, multivariate PDF is written in the form of
\begin{equation}\label{eq:JointPDF}
  f(x_1,\dots,x_D) = \left[\prod_{i=1}^D f_i(x_i)\right] c(F_1(x_1),\dots, F_D(x_D)).
\end{equation}
$f_i$ is the marginal PDF of variable $x_i$.
The last factor $c$, a multivariate function of marginal cumulative distribution functions (CDF) $F_i(x_i)$, 
stands for copula density, which designs the correlation between features. 
The copula density is set unity in the special case where variables are independent, but not in general.
In this form of PDF, one can select appropriate tools for the modeling marginals and copula density separately.
One typical choice is semiparametric copula, in which one uses a flexible non-parametric models to marginal distributions, 
and use a parametric model to copula density.
Gaussian copula is a simple member of parametric copula family, which can cope with correlations in
high ($D\geq 3$) dimension with simple structure based on multivariate normal distribution.
The copula density of Gaussian copula is defined as
\begin{align}\label{eq:copuladensity}
  c(F_1(x_1),&\dots, F_D(x_D)) = \nonumber\\
    & \left.\frac{\phi_D (\omega_1,\dots, \omega_D | \bm{0},\Sigma )}{\prod_{i=1}^D \phi_1(\omega_i)} \right|_{\omega_i = \Phi^{-1}\circ F_i(x_i)}
\end{align}
where $\phi_1$ and $\Phi$ are the PDF and CDF of one dimensional standard normal distribution, respectively.
$\phi_D $ is the PDF of $D$ dimensional multivariate normal distribution with zero mean vector 
and covariance matrix $\Sigma$, and $F_i$ is the CDF of $x_i$. 
Nonlinear monotonically non-decreasing transformations of probability variables $x_i\to \omega_i = \Phi^{-1}\circ F_i (x_i)$ 
are performed in which information of marginal distributions are eliminated and 
all the variables are forced to follow the standard normals $\omega_i \sim \mathcal{N}(0,1)$.
This transformation to normal variable is possible when the distribution $F_i$ is continuous on its support.
The correlation among variables are explained in terms of multivariate normal distribution 
with covariance $\Sigma$, after those nonlinear transformations. 
Mean vector of $\phi_D $ is zero and diagonal elements of $\Sigma$ are 1 since 
all the marginals are standard normal. 
The factors in denominator in eq.\eqref{eq:copuladensity} are in order for the integral of PDF to be unity.
As the probability measure changes under the transformation as
$f_i(x_i)\mathrm{d}x_i = \phi_1(\omega_i)\mathrm{d}\omega_i $, 
the normalization is ensured, 
$\int \mathrm{d}x_1\dots\mathrm{d}x_D f= \int \mathrm{d}\omega_1\dots\mathrm{d}\omega_D \phi_D =1$.
The estimation of $\Sigma$ is usually done by the maximum likelihood estimation (MLE) in $\omega$-space, 
in which empirical covariance matrix of observed $\omega$s is used for the estimator for $\Sigma$.

One well-known caveat in use of copulas is the handling of tied data~\cite{kojadinovic17some,Li20CopulaTie}. 
If many data points take same value, the transformation $\Phi^{-1}\circ F_i$ becomes discontinuous 
showing jump at the tied point, therefore transformed value $\omega_i$ no longer follows $\mathcal{N}(0,1)$.
Since zero inflation is one of the tied data, 
direct application of copula may results in a biased parameter estimation and poor fit.
In our second model, we demonstrate the discontinuity of transformation $\Phi^{-1}\circ F_i$, 
appropriate marginal distribution of $\omega_i$, 
and develop a new type of copula to handle the tied data.

\subsection{Mechanisms of Zero Inflations}
To better understand the statistical properties of zero-inflated data,
one needs to specify its mechanisms.
Here we show two mechanisms of zero inflation
in connection with the concepts in missing data analysis~\cite{howell07missing, MissingReview, little2019statistical}.

In the first scenario, zero-inflation is simply a lack of information, namely, 
zero-valued components of data are assumed to be obtained completely at random.
We assume that first we have $D$ variate positive-valued data, 
then a $D$ variate binary mask is applied on it, we eventually have $D$ variate zero-inflated data.
The mask, $D$-variate correlated Bernoulli type distribution, may be drawn for example from
restricted Boltzmann machines (RBM) ~\cite{Freund1991RBM1,Carreira2005RBM2},
which does not correlate to the original positive-valued data. 
We name this scenario as zero inflation completely at random (ZICAR)
after the similar concept in missing data analysis, missing completely at random (MCAR)\cite{little2019statistical}.
This scenario of zero inflation may happen when missing values in MCAR type are filled with zero values.

The other scenario is rather natural.
When we have zero-valued data, we can sometimes understand that there was actually 
a value significantly smaller than certain threshold.
The threshold may be determined by the resolution of 
measuring equipment or rounding rules in data accumulation process, for instances.
In this case, MCAR type missing scenario does not apply since the probability of zero occurrence 
strongly depends on the value supposed to be obtained. 
The situation is a special case of missing at random (MAR)\cite{little2019statistical} in that
zero occurrence happens deterministically.
We name this scenario as zero inflation by thresholding (ZIBT).

In subsequent sections, we provide two density estimation models corresponding to
ZICAR and ZIBT types, respectively.

\section{Proposal 1: Density Estimation for ZICAR Case}
As described, difficulties in density estimation for multivariate zero-inflated data 
lie in the exponentially large number of subspaces and the difference in dimensionality of subspaces.
Since subspace-wise estimation is exponentially complicated and 
unreliable due to the small number of data points in each subspace, 
we somehow need to estimate one multimodal $D$-variate density
by using data points as much as possible.

Our basic strategy behind our two models proposed in this and next sections are
to design the density model so that 
all the model parameters, defined in $D$ dimension, 
can be estimated in low dimension.
As will be demonstrated, marginalizability of copula-based density models ensures that parameters in $D$ dimensional density
directly appear in low dimensional marginalized density, 
hence we can estimate subset of high-dimensional parameters in low dimension.
When marginalized to low, namely two dimension, $2^D$ subspaces are projected onto only $2^2=4$ subspaces,
and we can fully utilize almost all of the data points for the estimation of parameters.

\subsection{Density Model}
In our first model, being specialized for ZICAR type scenario, 
we assume the $D$ dimensional positive-valued parent distribution, 
denoted by $g_D(x_1,\dots,x_D)$ for its PDF,
in which no zero inflation is included. 
The observed data with zero-inflation is assumed to be generated from the application of 
$D$ dimensional binary mask to $g_D$ as described in our definition of ZICAR.
We use Gaussian copula model to decompose $g_D$ into marginals and copula density as 
\begin{align}
  & g_D(x_1,\dots,x_D) \nonumber\\
  &= \left[\prod_{i=1}^D g_{\{i\} }(x_i)\right] 
    \left.\frac{\phi_D (\omega_1,\dots, \omega_D | \bm{0},\Sigma )}{\prod_{i=1}^D \phi_1(\omega_i)} \right|_{\omega_i = \Phi^{-1}\circ G_i(x_i)}
\end{align}
with $g_{\{i\} }(x_i)$ being the marginal PDF and $G_i$ being its CDF.
Similar to multivariate normal distribution, Gaussian copula model has a marginalizability,
a key feature repeatedly utilized in our proposal.
With the help of the formula for change of variable
$f_i(x_i)\mathrm{d}x_i = \phi_1(\omega_i)\mathrm{d}\omega_i $ and 
marginalizability of $\phi_D$, the PDF $g_D$ can easily be marginalized into arbitrary subset $(S\subseteq D)$ of variables as
\begin{align}\label{eq:marginalizability}
  g_{S}(\bm{x}_S) &= \int \mathrm{d} \bm{x}_{\bar{S}} \ g_D(x_1,\dots,x_D) \nonumber\\
  &= \left[\prod_{i\in S} g_{\iset}(x_i)\right] 
    \left.\frac{\phi_{|S|} (\bm{\omega}_S| \bm{0},\Sigma_S )}{\prod_{i\in S} \phi_1(\omega_i)} \right|_{\omega_i = \Phi^{-1}\circ G_i(x_i)}
\end{align}
with $S\subseteq \{1,\dots,D\}$, $\bar{S} = \{1,\dots,D\}\backslash S$ , $\bm{x}_S = \{x_i|i\in S\}$
and $\int \mathrm{d} \bm{x}_{\bar{S}} = \prod_{i\in\bar{S}} \int_0^{\infty} \mathrm{d}x_i $ .
In this paper, the integral measures are written in front of the integrand to clarify the ranges of integrals.
The covariance parameter $\Sigma_S = \{\Sigma_{ij}\}_{i\in S, j\in S}$ is the square submatrix of original $\Sigma$ 
restoring only dimensions in $S$. 
To simplify notations, singleton $\{i\}$ is sometimes written as $i$ 
and the full feature set $\{1,\dots,D\}$ is abbreviated by $D$. 
We are here ready to propose our first model, 
in which the total joint PDF of zero-inflated data after masking we express
\begin{equation}\label{eq:totalmodelZICAR}
  f_D(x_1,\dots,x_D) =  \sum_{S\subseteq D} q_S \delta (\bm{x}_{\bar{S}}) g_S(\bm{x}_S)
\end{equation}
where $\delta (\bm{x}_{\bar{S}}) \equiv \prod_{i\in \bar{S}} \delta (x_i)$ defines the $|S|$ dimensional subspaces.
$q_S \equiv P(\bm{x}_{\bar{S}}=0, \bm{x}_S>0)$ is the mask distribution dealing with the correlation of zero occurrences.
For the simplest case in which zero occurrences are mutually independent, the mask follows Bernoulli distributions, 
$q_S = \left[ \prod_{i\in \bar{S}}q_i \right] \left[ \prod_{j\in S}(1-q_j) \right]$ with marginal zero occurrence probabilities $q_i$.

\begin{figure}[!t]
  \centering
  \includegraphics[width=80mm]{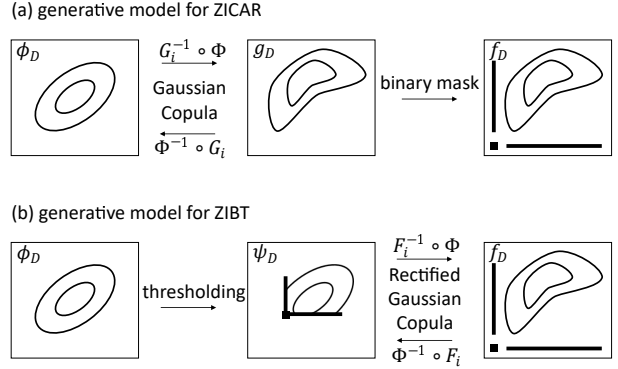}
  \caption{Summary of our two proposals in terms of generative models.
           Contours of distribution are schematically depicted. 
           Dots and straight lines indicate zero-valued data.          
          (a) For ZICAR model. Gaussian copula is used to express the parent distribution with its PDF $g_D$. 
          Multivariate zero-inflated distribution are assumed to be generated by $D$-variate binary masking to $g_D$.
          (b) For ZIBT model. Observed zero-inflated distribution are directly modeled by the 
          rectified Gaussian copula that is newly proposed in the present paper.
          Rectified Gaussian distribution with its PDF $\psi_D$, 
          a key building block in rectified Gaussian copula, 
          is generated by the thresholding to the multivariate normal distribution.
          The origin of zero inflation is the thresholding in this model.
          }
  \label{fig:GenerativeModels}
  \end{figure}

The overall generative process of our first model is summarized in Fig.~\ref{fig:GenerativeModels}(a).
The parent distribution with PDF $g_D$ is Gaussian copula model (middle panel) generated from 
dimension-wise nonlinear transformation to the multivariate normal distribution (left panel).
The total PDF $f_D$ (right panel) is then generated by applying multivariate binary mask to $g_D$.

\subsection{Estimation of Parameters}
Parameters to be estimated from observed zero-inflated data in eq.~\eqref{eq:totalmodelZICAR} 
are the mask distribution $q_S$, marginal distribution $g_{\iset}$, 
and covariance matrix in Gaussian copula $\Sigma$.
To estimate the mask distribution, 
one can first convert training data into binary matrix by the indicator function $I(x>0)$
in which positive values are replaced by one. 
Then one can use conventional estimation methods like RBM.

The estimation of marginal of parent distribution becomes apparent 
when the total PDF in eq.\eqref{eq:totalmodelZICAR} is marginalized into univariate.
The integral over $\bm{x}_{D\backslash \iset }$ can be executed by means of the marginalizability 
shown in eq.~\eqref{eq:marginalizability} and property of delta function as 
\begin{align}\label{eq:marginalPDF}
  f_{\iset}(x_i) &= \sum_{S\subseteq D} q_S \int \mathrm{d}\bm{x}_{D\backslash \iset} 
    \delta (\bm{x}_{\bar{S}}) g_S(\bm{x}_S) \nonumber \\
  &=  \sum_{S\subseteq D, i\notin S} q_S  \delta (x_i) 
    + \sum_{S\subseteq D, i\in S}    q_S g_\iset (x_i) \nonumber \\
  &=  q_{\iset}^{\iset} \delta(x_i) 
    + q_{\emptyset}^{\iset} g_\iset (x_i)\
\end{align}
with $q_{\iset}^{\iset} = P(x_i=0) = \sum_{S\subseteq D\backslash \iset} q_{S}$
and $q_{\emptyset}^{\iset} = 1- q_{\iset}^{\iset} = P(x_i>0) = \sum_{S\subseteq D\backslash \iset} q_{S\cup\iset}$ 
being the marginal mask distribution.
As the expression is completely parallel to eq.~\eqref{eq:pdf1dim}, 
marginal of parent distribution $g_\iset$ 
can be estimated from training data with $x_i>0$ using conventional methods like KDE.

The estimation of covariance matrix of Gaussian copula $\Sigma$ is slightly complex.
Most naively, $\Sigma$ can be estimated from training data with $x_i>0$ for all $i \in D$ with the 
empirical covariance matrix after transformation $\Phi^{-1}\circ G_i(x_i)$.
However, since the number of data points with all elements being positive is exponentially small in high dimension,
the estimation becomes unstable. 
Here we show an estimation method of $\Sigma$ which fully utilize almost all of the training data.
Similarly to eq.~\eqref{eq:marginalPDF}, the total PDF in eq.\eqref{eq:totalmodelZICAR} 
can be marginalized into bivariate as 
\begin{align}
  f_{\ijset}(x_i,x_j) = q_{\ijset}^{\ijset} \delta(x_i)\delta(x_j) &+ q_{\iset}^{\ijset} \delta(x_i)g_{\jset}(x_j) \nonumber\\
  + q_{\jset}^{\ijset} \delta(x_j)g_{\iset}(x_i) &+  q_{\emptyset}^{\ijset} g_{\ijset}(x_i, x_j) 
\end{align}
with $q_{S}^{\ijset} = P(\bm{x}_S=0, \bm{x}_{\ijset\backslash S}>0)$ for $S\subseteq\ijset$. 
In this form only the last term contains information of $\Sigma$, more specifically $\Sigma_{i,j}$.
Then, MLE of $\Sigma_{i,j}$ is the empirical covariance
between $\omega_i$ and $\omega_j$, 
in which only data points with $x_i>0$ and $x_j>0$ are used to compute the covariance.
It is worth noticing that one of the parameters in full $D$ dimensional model, namely $\Sigma_{i,j}$,
directly appears in $2$ dimensional marginalized distribution and hence can be estimated from $2$ dimensional expression.
Stability of estimates is expected to be enhanced with this method
since many data points in one quadrant can be used for the estimation.
Repeating this pair-wise estimations of $\Sigma_{i,j}$ for all pairs $1\leq i<j\leq D$,
all the components in $\Sigma$ can be estimated.

\subsection{Computation of Likelihood}
In order to use the estimated result to other applications such as unsupervised anomaly detections, 
the computation of likelihood function is needed. 
However as the total PDF in eq.\eqref{eq:totalmodelZICAR} in our hand contains Dirac's delta functions, 
value of PDF diverges. In this subsection we show a well-known treatment to handle the divergence of PDF
and to calculate finite values of likelihood.

This kind of divergence of PDF always appear in mixed random variable.
As well known for example in Tobit model~\cite{tobin58estimation}
and censored data in the field of survival analysis~\cite{kleinbaum12survival},
likelihood of discrete point in mixed random variable,
should be the value of the probability mass.
So, in our case, likelihood value corresponding to the PDF in eq.\eqref{eq:totalmodelZICAR} is written in the form of 
\begin{equation}
  L_D(x_1,\dots,x_D) =  \sum_{S\subseteq D} q_S I(\bm{x}_{\bar{S}}=0, \bm{x}_{S}>0) g_S(\bm{x}_S),
\end{equation}
which has non-diverging value for any data. Especially, for data with $x_{\bar{S}}=0, x_{S}>0$, 
likelihood is calculated as 
\begin{equation}
  L_D(x_1,\dots,x_D, \bm{x}_{\bar{S}}=0, \bm{x}_{S}>0 ) = q_S g_S(\bm{x}_S).
\end{equation}
This replacement of delta function by indicator function 
will be again used in our second model,
in which we present MLE of parameters using similarly constructed likelihood function.

\subsection{Rescaling of Variable}
Since the PDF of zero-inflated continuous data has different dimensionality for each subspace,
profile of likelihood function depends on the scaling of input variable.
This fact is clear even in one dimension. The likelihood function constructed from PDF of eq.\eqref{eq:pdf1dim} is
\begin{equation}
  L(x) =  qI(x=0) + (1-q)I(x>0) \tilde{f}(x).
\end{equation}
Not only the likelihood value itself, but also a balance between two terms 
is dependent on the scaling of variable $x$, due to the difference in dimensionality of two subspaces.
The different dependency on scaling is the build-in nature of mixed random variables,
which as well suffers our models when applied for example to unsupervised anomaly detection.
Here we propose one heuristic method to fully suppress the problems originating from the scaling of variable.
In this method, we redefine the scales of variables so that likelihood value after scaling satisfies desired property.
Firstly, we use data with original scales to estimate the marginal distributions as described above
and calculate the marginal likelihood.
Value of log marginal likelihood is $\log q_{\iset}^{\iset}$ for data with $x_i=0$
and $\log \left( 1- q_{\iset}^{\iset} \right) + \log g_{\iset}(x_i) $ for $x_i>0$,
in which $q_{\iset}^{\iset}$ is the estimated marginal zero occurrence rate. 
We here require the average of log marginal likelihood for data with $x_i>0$ to be $\log(1-q_{\iset}^{\iset})$. 
The variable $x_i$ then should be rescaled by $x_i\to x_i/b_i$ with 
\begin{equation}
  b_i = \exp \left[ -\frac{1}{N_i}\sum_{k, x^k_i>0} \log g_{\iset}(x^k_i) \right]
\end{equation}
where $N_i$ is the number of training data with $x^k_i>0$ and $x^k_i$ is the $i$-th component of $k$-th data. 
After this rescaling trick, we estimate marginal distributions again to get 
well-scaled final estimation results with desired property of marginal log likelihood functions.
This rescaling is also applicable to the second model we proposed in the next section.

\section{Proposal 2: Density Estimation for ZIBT Case}
In our first proposal, the zero inflation is assumed to happen in completely-at-random manner,
in which observed zero means just a lack of information and independent of the observed positive values.
This assumption might be sometimes unnatural since observed zero often indicates the tiny value smaller than certain threshold,
here we propose the second density model which assumes the ZIBT type of zero inflation.

The generative process assumed in our second model is depicted in Fig.~\ref{fig:GenerativeModels}(b).
In this model, the multivariate zero-inflated data (right panel) is directly modeled by copula. 
Application of Gaussian copula to zero-inflated data naturally needs
a use of rectified Gaussian distribution (middle panel), being generated by thresholding 
on the multivariate normal distribution (left panel), as described in detail in subsequent subsections.

\subsection{Rectified Gaussian Distribution}
In this model, our final joint PDF is directly decomposed by copula and has the form parallel to eq.\eqref{eq:JointPDF}.
However, as the nonlinear transformation becomes discontinuous for zero-inflated data, resultant variable $\omega_i = \Phi^{-1}\circ F_i(x_i)$
does not follow standard normal distribution and application of Gaussian copula results in poor fit to the data.
In order to show how to apply copula to this situation, 
we first illustrate the marginal and joint distribution of $\omega_i$ for zero-inflated variable.

The CDF $F_i(x_i)$ is formally defined by $P(X_i \leq x_i)$ with $X_i$ being the stochastic variable 
associated with its observation $x_i$. 
Then, inflated many zero data with $x_i=0$ are mapped together onto
$\Phi^{-1}\circ F_i(0) = \Phi^{-1}(q_i)$ where $q_i = F_i(0) = P(X_i\leq 0) = P(X_i=0)$ is the 
probability of zero occurrence in $X_i$.
On the other hand, as infinitesimal negative point $-0$ is mapped to 
$\Phi^{-1}\circ F_i(-0) = \Phi^{-1}\left( P(X_i \leq -0)\right) = \Phi^{-1}(0)=-\infty$, 
the transformation $\Phi^{-1}\circ F_i$ is discontinuous at zero. 
Transformed variable $\omega_i$ no longer follows $\mathcal{N}(0,1)$ due to the discontinuity.
For positive-valued data $x_i>0$, the transformation is $\omega_i = \Phi^{-1}\left(q_i + (1-q_i)\tilde{F}_i(x_i)\right)$ 
with $\tilde{F}_i(x_i) = \int_{-\infty}^{x_i} \mathrm{d}x \tilde{f}_i(x)$ being marginal CDF of the positive data distribution.
The density of transformed variable $\omega_i$ is then
$\left|\frac{\mathrm{d}x_i}{\mathrm{d}\omega_i}\right|(1-q_i)\tilde{f}_i =\phi_1(\omega_i)$,
hence positive data are mapped to the standard normal variable as if there were no zero inflations.
The marginal PDF of $\omega_i$ is then, instead of $\phi_1(\omega_i)$,
\begin{equation}\label{eq:RGD1dim}
  \psi_1(\omega_i|a_i) = q_i\delta(\omega_i-a_i) + I(\omega_i>a_i)\phi_1(\omega_i)
\end{equation}
with $a_i = \Phi^{-1}(q_i)$ being a threshold parameter of distribution.
Equivalently, $a_i$ is determined by the condition $q_i = \int_{-\infty}^{a_i} \mathrm{d} \omega \ \phi_1 (\omega)$. 
Although The PDF $\psi_1$ is similar to PDF of standard normal $\phi_1$,
the difference is that, the density fraction of $\phi_1$ below the threshold $a_i$ is integrated and gathered at $a_i$ resulting in 
a probability mass and expressed by a delta function.
Such distribution arising from thresholding are known as left-censored distribution~\cite{kleinbaum12survival}.
More specifically, the distribution generated from the thresholding of normal variable, 
is known to be a rectified Gaussian distribution (RGD)~\cite{Socci97RGD,Wu19RGDrelu}
and well studied for example in the field of factor analysis~\cite{harva07variational}.

Within this paper, we call the distribution of $\omega_i$ as standard RGD in that 
the location and scale of Gaussian component is $0$ and $1$, respectively.
Note that the mean and standard deviation of the distribution are different from the location and the scale.
The easiest generative model for standard RGD is written as
$\omega_i = \max\{a_i, \nu_i\}$ with $\nu_i\sim\mathcal{N}(0,1)$.
We here introduce rectifying operator
\begin{equation}
  \hat{R}_i(a_i) = \delta(\omega_i-a_i)\int_{-\infty}^{a_i} \mathrm{d} \nu_i + \int_{a_i}^{\infty} \mathrm{d} \nu_i\delta(\omega_i-\nu_i)
\end{equation}
where integrals are to be operated to the function placed in the right of the operator.
the marginal PDF~\eqref{eq:RGD1dim} can then be simplified as
\begin{equation}
  \psi_1(\omega_i|a_i) = \hat{R_i}(a_i) \phi_1( \nu_i).
\end{equation}

The extension of standard RGD to higher dimension is straightforward.
As for its generative model, 
we first draw a sample from $D$ variate normal distribution $(\nu_1,\dots,\nu_D)\sim\mathcal{N}(\bm{0},\Sigma)$,
then apply dimension-wise thresholding $\omega_i = \max\{a_i, \nu_i\}$,
then finally get a sample from multivariate standard RGD. 
As in the first model, $\Sigma$ has unit diagonal elements and zero location vector
since its all the marginals are standard RGD. 
The PDF of $D$ dimensional joint distribution then reads,
\begin{equation}\label{eq:PDFofRGD}
  \psi_D(\omega_1,\dots,\omega_D|\Sigma, \bm{a}) = \left[ \prod_{i=1}^D \hat{R}_i(a_i)\right] \phi_D(\nu_1,\dots,\nu_D|\Sigma)
\end{equation}
with $\bm{a}$ being the vector of thresholds $a_i$. As each operator acts only on one variable, 
they are commutative to each other.
A low dimensional example will be shown later in eq.\eqref{eq:PDFof2dimRGD}.

A key property of RGD is its marginalizability. The marginalization of high ($D$) dimensional RGD 
to any subset of features $S\subseteq D$ are again the RGD in the reduced subspace $S$.
This is apparent from the generative model and similar marginalizability of multivariate normal distribution.
The covariance in the reduced subspace is determined by $\{\Sigma_{i,j}\}_{i\in S, j \in S}$.
This property will be utilized in a future subsection to construct MLE.
Note here that the conditional distribution of RGD is not RGD, unlike the multivariate normal distribution.

\subsection{Density Model Using Rectified Gaussian Copula}
In the second model, we propose a new kind of copula associated with the multivariate RGD.
Firstly, we propose to roughly define our copula density as
\begin{align}
  c_{RGD}(F_1(x_1),&\dots, F_D(x_D)) = \nonumber\\
    & \left.\frac{\psi_D (\omega_1,\dots, \omega_D | \Sigma,\bm{a} )}{\prod_{i=1}^D \psi_1(\omega_i | a_i)} \right|_{\omega_i = \Phi^{-1}\circ F_i(x_i)}
\end{align}
and we name it rectified Gaussian copula.
Rectified nature of zero-inflated data after the transformation is directly modeled.
We can confirm that delta functions in the denominator are always canceled by those in numerator 
for any zero-occurrence patterns $S\subseteq D$.
Therefore our copula density is formally defined for each of subspace, by the reduction of fraction.
For subspace such that $\bm{x}_{\bar{S}}=0, \bm{x}_S>0$, the copula density is
\begin{align}\label{eq:RGDcopula_density}
  c_{RGD}(&F_1(x_1),\dots, F_D(x_D) ) = \nonumber\\
  & \left.\frac{\int\mathrm{d}\bm{\nu}_{\bar{S}} \phi_D(\bm{\nu}_{\bar{S}}, \bm{\omega}_{S}|\Sigma)}
  {\left[\prod_{i \in \bar{S}} \Phi(a_i) \right]\left[\prod_{j \in S}\phi_1(\omega_j)\right]}\right|_{\omega_i = \Phi^{-1}\circ F_i(x_i)}
\end{align}
in which $\int\mathrm{d}\bm{\nu}_{\bar{S}}$ 
is the abbreviation for $\prod_{i \in \bar{S}}\int_{-\infty}^{a_i} \mathrm{d} \nu_i$

As the marginals of original variables $x_i$ are expressed as in~\eqref{eq:pdf1dim}, 
the overall joint PDF of our second model is then
\begin{align}\label{eq:totalmodelZIBT}
    f_D(x_1,&\dots,x_D) = \nonumber\\
    & \left[\prod_{i=1}^D \left( q_i\delta(x_i) + (1-q_i) \tilde{f}_i(x_i) \right)\right] \times c_{RGD}.
\end{align}
In this form, density for $2^D$ subspaces are expressed all at once 
by $2^D$ terms after expanding the product.

\subsection{Estimation of Parameters}
In this subsection, we show methods to estimate parameters $q_i, \tilde{f}_i, a_i, 
\Sigma$ in our rectified Gaussian copula model.
Parameters in marginal distributions are easy to estimate. 
The marginal zero occurrence rate $q_i$ can be estimated from empirical zero value rate of $x_i$
and $\tilde{f}_i$ can be estimated again for example by KDE for data with $x_i>0$ only. 
The estimation of threshold parameter in rectified Gaussian copula density $a_i$ is apparent
from the relation $a_i = \Phi^{-1}(q_i)$. 
Estimation of the correlation parameter $\Sigma$ needs sophisticated analysis.

In our model, the total PDF of the form in eq.~\eqref{eq:totalmodelZIBT} assumes that
the distribution after transformation $\omega_i = \Phi^{-1}\circ F_i(x_i)$ is RGD 
with covariance $\Sigma$ and thresholds $\bm{a}$.
The problem here is to estimate parameters in multivariate RGD from observed data $\omega$,
here we show pair-wise MLE of the parameters. 
The marginalized PDF of RGD in 2 dimension is obtained from eq.~\eqref{eq:PDFofRGD} with the help of 
marginalizability of RGD as 
\begin{align}\label{eq:PDFof2dimRGD}
  &\psi_2 (\omega_i, \omega_j | \Sigma_{\ijset},\bm{a}_{\ijset} ) \nonumber\\
  &= \delta(\omega_i-a_i)\delta(\omega_j-a_j)\int_{-\infty}^{a_i} \mathrm{d} \nu_i\int_{-\infty}^{a_j} \mathrm{d} \nu_j \phi_2(\nu_i,\nu_j|\Sigma_{\ijset})\nonumber\\
  &+\delta(\omega_i-a_i)\int_{-\infty}^{a_i} \mathrm{d} \nu_i  \phi_2(\nu_i,\omega_j|\Sigma_{\ijset})\nonumber\\
  &+\delta(\omega_j-a_j)\int_{-\infty}^{a_j} \mathrm{d} \nu_j  \phi_2(\omega_i,\nu_j|\Sigma_{\ijset})\nonumber\\
  &+\phi_2(\omega_i,\omega_j|\Sigma_{\ijset})
  \end{align}
with $\bm{a}_{\ijset}=(a_i,a_j)$ and $\Sigma_{\ijset}= 
\left(\begin{matrix}   1 & \Sigma_{i,j} \\   \Sigma_{i,j} & 1 \end{matrix}\right)$.
As in the first model, $\Sigma_{i,j}$, one of the parameters in the full model, directly appears in the 2 dimensional PDF,
enabling us to estimate using many data points.
However, unlike the first model, as the 2 dimensional distribution is RGD, not a normal distribution, 
empirical covariance between $\omega_i$ and $\omega_j$ is no longer an MLE.
The true likelihood for $1$ data here is eq.~\eqref{eq:PDFof2dimRGD} with replacement of delta functions
by the indicator functions, namely,
\begin{align}\label{eq:likelihood2D}
&L(\omega_i, \omega_j | \Sigma_{\ijset},\bm{a}_{\ijset} ) \nonumber\\
&=\left\{
  \begin{array}{ll}
    \int_{-\infty}^{a_i} \mathrm{d} \nu_i\int_{-\infty}^{a_j} \mathrm{d} \nu_j \phi_2(\nu_i,\nu_j|\Sigma_{\ijset}) & (x_i=x_j=0) \\
    \int_{-\infty}^{a_i} \mathrm{d} \nu_i  \phi_2(\nu_i,\omega_j|\Sigma_{\ijset}) & (x_i=0, x_j>0) \\
    \int_{-\infty}^{a_j} \mathrm{d} \nu_j  \phi_2(\omega_i,\nu_j|\Sigma_{\ijset}) & (x_i>0, x_j=0)  \\
    \phi_2(\omega_i,\omega_j|\Sigma_{\ijset}) & (x_i>0, x_j>0)
  \end{array}
\right. \nonumber \\
&=\left\{
  \begin{array}{ll}
    \Phi_2(a_i,a_j|\Sigma_{\ijset}) & (x_i=x_j=0) \\
    \phi_1(\omega_j)\Phi(a_i|\Sigma_{i,j}\omega_j , 1-\Sigma_{i,j}^2) & (x_i=0, x_j>0) \\
    \phi_1(\omega_i)\Phi(a_j|\Sigma_{i,j}\omega_i , 1-\Sigma_{i,j}^2) & (x_i>0, x_j=0)  \\
    \phi_2(\omega_i,\omega_j|\Sigma_{\ijset}) & (x_i>0, x_j>0)
  \end{array}
  \right.
\end{align}
with $\Phi_2$ being the CDF of $2$ dimensional normal distribution. 
The log likelihood is given by the logarithms of eq.~\eqref{eq:likelihood2D} summed over all the training data. 
The MLE is then obtained by numerically maximizing the total log likelihood with 
respect to $\Sigma_{ij}$ in the open interval $(-1,1)$.
Repeating those pairwise estimations for all pairs, we can get full matrix estimate for $\Sigma$.

The estimation of $\Sigma$ developed here is a high dimensional extension 
of the method in literature~\cite{Li20CopulaTie},
in which general bivariate copula estimation method for generally tied data 
is proposed.
The marginalizability of RGD to 2 dimension makes the use of bivariate method possible, 
hence the $D$-dimensional copula parameters can be estimated.

The authors in literature \cite{Wu19RGDrelu} has developed an angle based pair-wise estimation method of $\Sigma$
in RGD with its error bounds, however the nonnegative assumption they made in the literature,
corresponding to $a_i<0$ (equivalently $q_i<1/2$) for all $i$ in our formulation, is sometimes violated.
They showed that the required number of data points to bound the estimation error below a fixed value
exponentially increases as $\Omega(\exp (a_i^2/2))$ when $a_i>0$ 
since the most of the samples fall outside the threshold then information is lost.
On the other hand in our estimation method in the second model, we fully utilize all the data 
including rectified ones, we expect more efficient estimation is realized even for $q_i>1/2$.
Mathematical supports of our method, such as error bounds and required sample size
are left for future study.

\subsection{Computation of Likelihood}
For the calculation of full likelihood value in $D$ dimension,
we first replace delta functions in PDF by indicator functions.
The process is a generalization of the derivation of eq.~\eqref{eq:likelihood2D} from eq.\eqref{eq:PDFof2dimRGD}.
The likelihood value is therefore,
\begin{align}\label{eq:likelihoodD}
  &L(x_1,\dots x_D , x_{\bar{S}}=0, x_{S}>0 ) \nonumber \\
  &= \left[\prod_{i \in \bar{S}} q_i \right] \left[\prod_{j \in S} (1-q_j) \tilde{f_j}(x_j)\right] c_{RGD}.
\end{align}

For the implementation,
integral over $\bm{\nu}_{\bar{S}}$ 
in rectified gaussian copula density $c_{RGD}$ defined in eq.~\eqref{eq:RGDcopula_density}
can be evaluated using libraries
capable to numerically compute the CDF of multivariate normal distribution of the form
\begin{align}
&\int\mathrm{d}\nu_{\bar{S}} \phi_D(\bm{\nu}_{\bar{S}}, \bm{\omega}_{S}|\Sigma) \nonumber\\ 
&= \phi_{|S|}(\bm{\omega}_{S}|\Sigma_{S}) \int\mathrm{d}\bm{\nu}_{\bar{S}} \phi_{D-|S|}(\bm{\nu}_{\bar{S}}|\bm{\mu}_{\bar{S}|S}, \Sigma_{\bar{S}|S}).
\end{align}
Conditional mean and covariance is determined by the conventional results for multivariate normal distribution as
$\bm{\mu}_{\bar{S}|S} = \Sigma_{\bar{S}S}\Sigma_S^{-1}\bm{\omega}_S$ and 
$\Sigma_{\bar{S}|S} = \Sigma_{\bar{S}} - \Sigma_{\bar{S}S} \Sigma_S^{-1} \Sigma_{\bar{S}S}^T$.
Depending on the algorithm for the evaluation of the CDF,
the computation of the likelihood may need a exponentially long time as $D$ increases.
Although one can use a well-designed algorithm for the calculation of CDF
of multivariate normal distribution, for example as in~\cite{gessner20integrals},
we here provide an approximate computation method of the likelihood,
in which we partially neglect the correlation among zero occurrences as
\begin{align}
  \int\mathrm{d}\bm{\nu}_{\bar{S}} \phi_D(\bm{\nu}_{\bar{S}}, \bm{\omega}_{S}|\Sigma) 
  &\approx \phi_{|S|}(\bm{\omega}_{S}|\Sigma_{S}) \int\mathrm{d}\bm{\nu}_{\bar{S}} 
  \prod_{i\in\bar{S}}\phi_1(\nu_i) \nonumber\\
  &= \phi_{|S|}(\bm{\omega}_{S}|\Sigma_{S}) \prod_{i\in\bar{S}}q_i .
  \end{align}
This approximation greatly reduces the computational cost 
by bypassing the high dimensional numerical integration 
and computation is clearly within polynomial time;
\begin{equation}\label{eq:approx}
  \tilde{c}_{RGD} 
  \approx \frac{\phi_{|S|}(\bm{\omega}_{S}|\Sigma_{S}) }{\prod_{j \in S}\phi_1(\omega_j)} 
  =  (2\pi)^{|S|/2}\phi_{|S|}(\bm{\omega}_{S}|\Sigma_{S}-I_S)
\end{equation}
with $I_S$ being the identity matrix of size $|S|$.
Although the approximated form of $\tilde{c}_{RGD}$ only consider the correlations among positive variables,
parameter $\Sigma$ is what we estimated using all the training data containing some zero elements.

We conclude the last two sections about our proposals by illustrating the 
differences and similarities of our two models.
First model is for ZICAR case in which zero inflations are completely 
at random regardless of the values which supposed to be obtained. 
It also can be applied to missing data with MCAR type scenario.
As the second model considers that the fictitious values behind the observed zeros are 
smaller than certain threshold, 
the latter model seems to be much more natural for realistic mechanism for zero inflation.
On the other hand, correlation between zero occurrences in different variable
is better captured in the first model 
since we can use arbitrary multivariate binary mask model such as RBM.
Correlation between zeros in the second model is only expressed through 
the thresholding of multivariate normal distribution:
\begin{align}
  &P(\bm{x}_{\bar{S}}=0, \bm{x}_S>0) \nonumber \\
  &= \int_{-0}^{+0} \mathrm{d} \bm{x}_{\bar{S}} \int_{+0}^{\infty} \mathrm{d} \bm{x}_S \  f_D(x_1,\dots,x_D ) \nonumber\\
  &= \left[\prod_{i\in\bar{S}} q_i \right] \int_{+0}^{\infty} \mathrm{d} \bm{x}_S \ 
    \left[\prod_{i\in S} (1-q_i)\tilde{f}_i(x_i)\right] \left.c_{RGD}\right|_{\bm{x}_{\bar{S}}=0, \bm{x}_S>0} \nonumber \\
  &= \int_{+0}^{\infty} \!\!\! \mathrm{d} \bm{x}_S \!\left[\prod_{i\in S} (1-q_i)\tilde{f}_i(x_i)\right] \!\!
    \left.\frac{\int\! \mathrm{d}\bm{\nu}_{\bar{S}} \phi_D(\bm{\nu}_{\bar{S}}, \bm{\omega}_{S}|\Sigma)}
    {\left[\prod_{j \in S}\phi_1(\omega_j)\right]}\right|_{\omega_i = \Phi^{-1}\circ F_i(x_i)} \nonumber \\
  &= \left[ \prod_{i\in S} \int_{a_i}^{\infty} \mathrm{d} \bm{\omega}_i \right]
    \left[ \prod_{j \in \bar{S}} \int_{-\infty}^{a_i} \mathrm{d} \nu_{\bar{j}}\right] 
    \phi_D(\bm{\nu}_{\bar{S}}, \bm{\omega}_{S}|\Sigma).
\end{align}
The probability of each zero occurence pattern is expressed by the density fraction of $\mathcal{N}(\bm{0},\Sigma)$
integrated over one orthant separated by thresholds $\bm{a}$.
In other words, it is equivalent to $D$ variate binary 
variables expressed by $I(\omega_i > a_i)$ with 
$(\omega_1,\dots,\omega_d)\sim \mathcal{N}(\bm{0},\Sigma)$.
This class of multivariate Bernoulli distribution, and its continuous relaxation, 
is discussed in ref.~\cite{wang20MBD}. As the degrees of freedom of the distribution 
is order of $D^2$, the expressive power is lower than the
sophisticated models such as RBM. 
Correlations among positive values, among zeros, and between them are all expressed 
in a single rectified Gaussian copula and entangled in the ZIBT model.
Both of our two models are capable of treating exponentially large number of subspaces
with different dimensionalities.
Estimation of parameters and the computation of the likelihood are 
both within polynomial time, when approximated likelihood is used in ZIBT model.

\section{Numerical Experiments}
We have performed a set of numerical experiments to show 
the performance of our density estimation models
using synthetic and real multivariate zero-inflated data.

We note that a widely used performance metric for density estimation, total log likelihood, 
is not appropriate for zero-inflated data due to the difference of dimensionality of subspaces.
For example, GMM model can earn infinitely large value of likelihood by the infinitesimal width 
of a Gaussian component located at a low dimensional subspace.
In our experiment, therefore, the performance is measured in a supervised anomaly detection setting~\cite{emmott15meta}.
Set of normal test data is drawn from given ground truth distribution or split from real dataset.
All the normal test data are duplicated and corrupted to generate abnormal data, then
the performance of anomaly detection task is measured.
The area under the receiver operating characteristic curve(AUC) 
value between the negative log likelihood of model output 
and the binary abnormal flag is used for the metric.
In the corruption process, the positive valued component $x_i$ is replaced by
an i.i.d. sample from the uniform distribution between $m_i$ and $M_i$. Zero-valued components are not corrupted.
The bounds $m_i$ and $M_i$ are defined by $1$ and $99$ percentile points of $x_i$ in the training data with $x_i>0$, respectively.
Note here that the task is getting easier for higher dimension $D$, 
since the data manifold becomes smaller compared to the hyperrectangles specified by $m_i$ and $M_i$. 

\begin{table*}[!t]
  \renewcommand{\arraystretch}{1.1}
  \caption{Averaged AUC values for each synthetic data with different model. 
  The highest values are indicated by underlines.}
  \label{tab1}
  \centering
  \begin{tabular}{clccccccccccc}\hline
    \multicolumn{4}{c}{}  & \multicolumn{4}{c}{ZICAR model} & \multicolumn{4}{|c}{ZIBT model}\\
    \cline{5-8}\cline{9-12}
    \multicolumn{2}{c}{}&   GMM    & KDE     & \multicolumn{1}{c}{full}   
    & \begin{tabular}{c} w/o \\ RBM \end{tabular}  
    & \begin{tabular}{c} w/o \\ MLE \end{tabular} 
    & \begin{tabular}{c} w/o \\ rescale \end{tabular} 
    & \multicolumn{1}{|c}{full} 
    & \begin{tabular}{c} w/ \\ approx. \end{tabular} 
    & \begin{tabular}{c} w/o \\ MLE \end{tabular}  
    & \begin{tabular}{c} w/o \\ rescale \end{tabular}  \\ \hline
               & $D=2 $ & 0.6502 & 0.6871 & \underline{0.8091} & 0.7983 & 0.7514 & 0.7888 & 0.7308 & 0.7353 & 0.7291 & 0.6868 \\
          ZICAR& $D=5 $ & 0.7460 & 0.8598 & \underline{0.9274} & 0.9207 & 0.8483 & 0.9145 & 0.8279 & 0.8332 & 0.8292 & 0.7768 \\
          DATA & $D=10$ & 0.8491 & 0.8663 & \underline{0.9534} & 0.9513 & 0.9126 & 0.9489 & 0.8995 & 0.9048 & 0.8992 & 0.8554 \\ 
               & $D=15$ & 0.8719 & 0.7849 & \underline{0.9795} & 0.9789 & 0.9408 & 0.9777 & 0.9310 & 0.9378 & 0.9332 & 0.8960 \\ \hline
               & $D=2 $ & 0.6921 & 0.7032 & 0.7359 & 0.7402 & 0.7337 & 0.7375 & \underline{0.7717} & 0.7587 & 0.7680 & 0.7550 \\
          ZIBT & $D=5 $ & 0.7732 & 0.8636 & 0.8165 & 0.8117 & 0.8069 & 0.8076 & \underline{0.9181} & 0.8742 & 0.8901 & 0.9081 \\
          DATA & $D=10$ & 0.8899 & 0.9211 & 0.9034 & 0.8915 & 0.9061 & 0.8985 & \underline{0.9792} & 0.9525 & 0.9581 & 0.9786 \\ 
               & $D=15$ & 0.9381 & 0.9112 & 0.9579 & 0.9480 & 0.9574 & 0.9448 & \underline{0.9959} & 0.9838 & 0.9901 & 0.9945 \\ \hline
      \end{tabular}
  \end{table*}

\subsection{Ablation Study in Synthetic Data}
First experiment is done by using synthetic data. 
We generate dataset from both the generative processes 
depicted in Fig.~\ref{fig:GenerativeModels}(a) and (b).

\subsubsection{Data Generation Process}

For ZICAR data, details of data generation processes are followings. 
First, we randomly initialize the ground truth distribution.
Parameters to be initialized are marginal distribution $g_{\iset}$, 
covariance parameter $\Sigma$, and binary mask distribution.
In stead of randomly initializing the marginal distribution or its CDF $G_i$, 
we sample random strictly increasing functions $h_i: (-\infty,\infty) \to (0,1)$
and use $h_i$ instead of $G_i^{-1}\circ \Phi$.
$h_i$ is generated by a weighted sum of sigmoid functions 
$h_i(x_i) = \sum_{j=1}^5 \pi_{i,j} \sigma(b_{i,j}( x_i-c_{i,j}))$ with $\pi_{i,j}$ being a weight 
drawn from the Dirichlet distribution with parameter $(1,1,1,1,1)$, 
$b_{i,j}$ drawn uniformly from $(0,2)$ and $c_{i,j}$ drawn uniformly from $(-5,5)$.
Covariance $\Sigma$ of multivariate normal is drawn from 
Wishart distribution with degrees of freedom $D$ 
and matrix-valued parameter $I_D$. $\Sigma$ is then normalized as 
$\Sigma_{ij} \to \Sigma_{ij} / \sqrt{| \Sigma_{ii}\Sigma_{jj}|}$ 
to ensure the diagonal elements to be unity.
Multivariate binary mask distribution is expressed by
RBM defined with $D$ visible units and $2^{D/2}$ hidden units.
Weights for the hidden layer and interactions are from $\mathcal{N}(0,0.1^2)$ and 
for visible units are from $\mathcal{N}(1,0.1^2)$.
Next, we draw samples from fixed ground truth distribution. 
Multivariate normal sample is drawn from $\mathcal{N}(\bm{0}, \Sigma)$ and random monotonic function $h_i$ is applied per element 
in order to make the samples follow the distribution depicted in the middle panel of Fig.~\ref{fig:GenerativeModels}(a).
The binary mask is drawn from the RBM and applied to the sample to get ZICAR data as depicted in the right panel of Fig.~\ref{fig:GenerativeModels}(a).

For ZIBT data, 
parameters to be initialized are 
marginal PDF for positive data $\tilde{f}_i$, 
marginal zero occurrence rate $q_i$,
threshold $a_i$, and covariance matrix $\Sigma$.
Covariance $\Sigma$ is generated in a same way as ZICAR case.
The marginal zero rate $q_i$ is sampled uniformly from $(0,0.5)$
and converted to thresholds; $a_i = \Phi^{-1}(q_i)$.
A discontinuous random non-decreasing function,
used as $F^{-1}\circ\Phi$ in Fig.~\ref{fig:GenerativeModels}(b), is sampled as follows.
Firstly, $M=10^4$ i.i.d. standard normal samples are drawn and random strictly increasing function,
same as in ZICAR case, is applied.
Next, $M$ i.i.d. binary samples, drawn from Bernoulli distribution with mean $1-q_i$,
are multiplied to the sample then 
$M$ samples from univariate random zero-inflated distribution are obtained.
Then a non-decreasing mapping from the samples to univariate standard RGD, 
used instead of $\Phi^{-1}\circ F_i$, can be constructed empirically. 
Finally we get discontinuous random non-decreasing function $h_i$, 
a sample of $F_i^{-1}\circ\Phi$, by inverting the constructed empirical mapping.
In data sampling process from fixed ground truth distribution,
multivariate normal sample is drawn from $\mathcal{N}(\bm{0}, \Sigma)$ 
and thresholded at each $a_i$ to generate multivariate RGD data 
as in the middle panel of Fig.~\ref{fig:GenerativeModels}(b).
Applying the discontinuous random monotonic functions $h_i$ constructed above in each element,
eventually we get a ZIBT data as in the right panel of Fig.~\ref{fig:GenerativeModels}(b).

For both cases, training and test data sizes are $10000$.
The half of the test data is normal data sampled from a ground truth distribution,
the rest $5000$ are abnormal test data generated by the corruption to the duplicated normal test data.
AUC scores are calculated for 15 trials with different random seeds and averaged values will be shown.

\subsubsection{Ablations}
We conduct an ablation study in order to test the effectiveness of each component of our models.
We use following suffixes to distinguish tested variants of each of two models;
\begin{itemize}\setlength{\leftskip}{15mm}
  \item[full: ] Full version of each model with all components used. 
              Marginal distribution of positive data is estimated by KDE, 
              and binary mask in ZICAR model is learned by RBM with $D$ visible and $2D$ hidden layers
  \item[w/o RBM: ] Binary mask is independent Bernoulli distribution, without using RBM. 
              This variant is only available for ZICAR model
  \item[w/ approx.: ] Likelihood function is computed using approximation in eq.~\eqref{eq:approx}. 
              This variant is only available for ZIBT model
  \item[w/o MLE: ] Estimation of $\Sigma$ is done by empirical covariance matrix 
              using all data including zero-valued ones
  \item[w/o rescale: ] The heuristic rescaling of the variables is not used
\end{itemize}
As baselines, we also measured the performances of conventional density estimation models, GMM and KDE.
All the hyper parameters are tuned using independently generated data from the initialized distributions before each of 15 trials.

\subsubsection{Results}
The measured average AUC values are summarized in table~\ref{tab1}.
For each data generation process, ZICAR data and ZIBT data, we set different dimension $D$.
The best scores are indicated by underlines.
We can clearly see that full models show best performances for all dimensions, 
and as expected, best model for ZICAR/ZIBT data is full ZICAR/ZIBT model, respectively,
which confirms that our two full models surely capture the ground truth distributions.
Even if we fail to choose an appropriate model for the zero inflation mechanisms of data,
e.g., if we mistakenly select ZIBT model for ZICAR data or vice versa,
the performance is still higher than baselines in most cases, 
reflecting the suitable handling of the different dimensionality of the zero-inflated data in our models.

For the models with ablations, the performances slightly degrades from the corresponding full model.
That strongly indicates all the components contribute to the model quality.
The performances of ablation models are still better than the baselines and 
improperly selected model, e.g., ZICAR model with ablations are better than full ZIBT model for ZICAR data.
An important lesson from those observations is that to select an 
appropriate model suitable for the mechanism of zero-inflation in data,
from ZICAR, ZIBT, or other future models, 
is more important than the complicated components inside the model.

The largest deterioration happens in w/o MLE type of ablation for ZICAR data and ZICAR model 
while degradation is small in the ZIBT data and ZIBT model.
To further inspect the effectiveness of MLE, we summarized the $L^2$ estimation error 
of covariance matrix, averaged over 15 trials, for each data and model in table~\ref{tab2}.
As it should be, estimation error is the smallest for appropriate full model with MLE (indicated by underlines).
The estimation error is large in ZICAR w/o MLE model for ZICAR data and small in the ZIBT counterpart, 
consistent with the largest deterioration of AUC in ZICAR w/o MLE model for ZICAR data.

\begin{table}[!t]
\renewcommand{\arraystretch}{1.1}
\caption{Averaged $L^2$ estimation error of covariance matrix for each data with different estimation methods. 
The smallest values are indicated by underlines.}
\label{tab2}
\centering
\begin{tabular}{clccccc}\hline
  \multicolumn{2}{c}{}  & \multicolumn{2}{c}{ZICAR model} & \multicolumn{2}{c}{ZIBT model}\\
  \cline{3-4}\cline{5-6}
  \multicolumn{2}{c}{}& full &\begin{tabular}{c} w/o \\ MLE \end{tabular} & full  &\begin{tabular}{c} w/o \\ MLE \end{tabular} \\ \hline
            & $D=2 $  &  \underline{0.005} & 0.866 & 0.641 & 0.524 \\
       ZICAR& $D=5 $  &  \underline{0.052} & 1.716 & 1.302 & 1.063 \\
       DATA & $D=10$  &  \underline{0.118} & 2.658 & 2.042 & 1.665 \\ 
            & $D=15$  &  \underline{0.195} & 3.363 & 2.537 & 2.066 \\ \hline
            & $D=2 $  &  0.190 & 0.112 & \underline{0.005} & 0.062 \\
       ZIBT & $D=5 $  &  0.848 & 0.326 & \underline{0.037} & 0.305 \\
       DATA & $D=10$  &  1.321 & 0.514 & \underline{0.096} & 0.421 \\ 
            & $D=15$  &  1.709 & 0.668 & \underline{0.148} & 0.528 \\ \hline
\end{tabular}
    
\end{table}

\subsection{Experiment For Real Zero-Inflated Data}
The second experiment uses a real world multivariate zero-inflated data.

\subsubsection{Data}
Open dataset of multivariate zero-inflated nonnegative continuous data
with significant amount of zero-valued data is quite rare.
UCI credit data~\cite{UCI, UCIcredit} contains 30000 records of information
such as credit limit, sex, education, marital status, history of payment, and so on 
of individuals in a credit card service in Taiwan.
The 12 variables $\textit{PAY\_AMT\{i\}}$ and $\textit{BILL\_AMT\{i\}}$ with $i=1,\dots,6$,
indicating paid amount and billed amount of the customer $i$ months before,
are the correlated zero-inflated continuous data with significant zero value rate
about $6\sim 24\%$ and right-skewed distribution.
We use those 12 variables in the present experiment.
Note, tiny amount of negative values exist and are replaced by zeros.
Randomly selected 21000 records are used as a training data,
while the rest 9000 records are for normal test data.
9000 records of abnormal test data are 
again generated by the corruption to the duplicated normal test data.
All the hyper parameters are tuned within the training data.
Averaged values of AUC over 15 random seeds are calculated.
In addition to the experiment for $D=12$ using all the zero-inflated variables
we performed a experiment for $D=2$ in which we only use $\textit{PAY\_AMT1}$ and $\textit{BILL\_AMT1}$,
to evaluate the performance in low dimensional case.

\subsubsection{Results}
Averaged AUC values are summarized in table~\ref{tab3}.
All of our models and its approximant exceed the baselines.
ZIBT full models shows the best performance both for $D=2$ and $D=12$.
The result implies that ZIBT type scenario is more appropriate than ZICAR for 
the mechanism of zero inflation in UCI credit data, 
being consistent with our intuition.
Namely, the zero-valued data are not just a lack of information but indicate smaller value than certain threshold.
The drop of the performance in the ZIBT w/approx. model is quite tiny,
despite the remarkably improved computational efficiency 
by bypassing the high dimensional numerical integration.

\begin{table}[!t]
\renewcommand{\arraystretch}{1.1}
\caption{Averaged AUC values for UCI credit data with different models. The largest values are indicated by underlines.}
\label{tab3}
\centering
\begin{tabular}{lccccc}\hline
  \multicolumn{3}{c}{}  
  & \multicolumn{1}{c}{\begin{tabular}{c} ZICAR \\ model \end{tabular} } 
  & \multicolumn{2}{c}{ ZIBT  model }\\
  \cline{4-4}\cline{5-6}
    & GMM    & KDE    & full   & full   & \begin{tabular}{c} w/\\approx.\end{tabular}\\ \hline
    $D=2 $  & 0.866 & 0.899 & 0.906 & \underline{0.917} & 0.916 \\
    $D=12$  & 0.953 & 0.953 & 0.966 & \underline{0.987} & 0.985 \\ \hline
  \end{tabular}
\end{table}

\section{Conclusion and Future Directions}

Density estimation of multivariate zero-inflated non-negative continuous data is challenging due to
exponentially large number of subspaces, different dimensionality of them, and diverging PDF.
In this paper, we discussed two types of scenarios of zero inflation, ZICAR and ZIBT,
and correspondingly presented two copula-based density estimation models, ZICAR model and ZIBT model.
Especially in the ZIBT model, we proposed a new variant of copula, 
rectified Gaussian copula to accommodate the tied data at zero
and developed an MLE for RGD.
In both models, the marginalizability of the density models 
enables to construct efficient parameter estimation methods
that fully utilize almost all of the training data.
Estimation of parameters and computation of likelihood are both within polynomial time.
Difficulties in density estimations of multivariate zero-inflated data, 
originating from exponentially large number of subspaces, 
different dimensionality of them, and diverging PDF,
are all well-handled with our models.

In the numerical experiment on artificially generated ZICAR and ZIBT data, 
we have shown that properly selected model for each of the zero-inflation scenarios exhibits the best fit to the data.
We have also confirmed all the components of our proposal, namely RBM, MLE, and rescaling trick, 
works positively on the performance of models.
Numerical experiment for real-world zero-inflated data has demonstrated the effectiveness 
of our models compared to the conventional models like GMM and KDE.
The best performance is governed by ZIBT model.
It indicates the ZIBT type scenario, in which zero-valued data are considered to be small value, not just a lack of information, 
is more appropriate for this data as we expect.

In the density estimation for multivariate zero-inflated data, 
one needs to capture all of 
correlation among positive values, 
correlation among zero occurrences, and
correlation between positive and zero.
Our ZICAR model cannot capture the third type of correlation
while in the ZIBT model, all types are simultaneously 
expressed in a single rectified Gaussian copula then entangled.
We may need more flexible and expressive modeling tools for multivariate zero-inflated data.

\section*{Acknowledgments}
The author would like to thank S. Matsumoto for valuable comments on the earlier draft of the manuscript,
and T. Yoshiba and M. Kazato for fruitful discussions.

\bibliographystyle{unsrt.bst}

\bibliography{main_submit_arxiv.bbl}

\end{document}